\documentclass[epsfig,12pt]{article}
\usepackage{epsfig}
\usepackage{graphicx}
\usepackage{amsmath}
\usepackage{array}
\usepackage{color}
\usepackage{bm}
\usepackage[colorlinks,citecolor = blue]{hyperref}

\usepackage[nottoc,notlof,notlot]{tocbibind} 
\usepackage[titles]{tocloft}

\newcommand {\s}{\vspace{.25in}}
\newcommand{\beq}{\begin{equation}}   
\newcommand{\eeq}{\end{equation}}
\newcommand{\beqn}{\begin{eqnarray}}   
\newcommand{\eeqn}{\end{eqnarray}}

\newcommand{\pt}{\partial}

\newcommand{\gsim}{\lower.7ex\hbox{$
\;\stackrel{\textstyle>}{\sim}\;$}}
\newcommand{\lsim}{\lower.7ex\hbox{$
\;\stackrel{\textstyle<}{\sim}\;$}}
\usepackage{tikz}
\usetikzlibrary{decorations.shapes}
\usepackage{caption}
\usepackage{subcaption}
\usepackage[compat=1.0.0]{tikz-feynman}
\usepackage{amssymb}
\usepackage[colorinlistoftodos]{todonotes}
\def\a{\alpha}

\def\b{\beta}
\def\c{\chi}

\def\D{\Delta}

\def\eps{\varepsilon}
\def\f{\frac}

\def\G{\Gamma}

\def\l{\left}

\def\mc{\mathcal}

\def\m{\mu}
\def\n{\nu}

\def\nn{\nonumber}

\def\p{\partial}
\def\vp{\varphi}

\def\r{\right}
\def\s{\sigma}
\def\t{\theta}

\def\Tr{\mathrm{Tr}}

\def\vp{\varphi}

\def\dst{\displaystyle}

\def\be{\begin{equation}}
\def\ee{\end{equation}}

\def\bea{\begin{eqnarray}}
\def\eea{\end{eqnarray}}

\def\ba{\begin{array}}
\def\ea{\end{array}}

\def\bc{\begin{center}}
\def\ec{\end{center}}

\def\bl{\begin{flushleft}}
\def\el{\end{flushleft}}

\def\br{\begin{flushright}}
\def\er{\end{flushright}}

\def\bi{\begin{itemize}}
\def\ei{\end{itemize}}

\def\bt{\begin{tabular}}
\def\et{\end{tabular}}

\newcommand{\REF}[1]{(\ref{#1})}

\RequirePackage{luatex85}
\begin{document}

\begin{flushright}
FTPI-MINN-23-09, UMN-TH-4215/23
\end{flushright}

\begin{center}
{\Large \bf Spectral Flow in Instanton Computations\\ and the \boldmath{$\b$} functions}

\vspace*{0.5cm}

 {Alexander~Monin,$^{1}$ Mikhail Shifman,$^{2}$ and Arkady Vainshtein$^{2,3}$}

\vspace*{0.5cm}

$^{1}${\it Department of Physics and Astronomy, \\
University of South Carolina, \\
Columbia SC 29208, USA}

\hspace*{0.3cm}

$^{2}${\it William I. Fine Theoretical Physics Institute, \\
University of Minnesota, \\
Minneapolis, MN 55455, USA}

\hspace*{0.3cm}

$^{3}${\it Kavli Institute for Theoretical Physics, \\
University of California, \\
Santa Barbara, CA 93106, USA}

\vspace*{0.5cm}
\texttt{\small amonin@mailbox.sc.edu}  \\
\texttt{\small shifman@umn.edu}  \\
\texttt{\small vainshte@umn.edu} \\

\vspace{1cm}

\end{center}

\begin{abstract}

We discuss various differences in the instanton-based calculations of the $\beta$ functions in theories such as Yang-Mills and $\mathbb{CP}(N\!-\!1)$ on one hand, and $\lambda\phi^4$ theory with Symanzik's sign-reversed prescription for the coupling constant $\lambda$ on the other hand. Although the aforementioned theories are asymptotically free, in the first two theories, instantons are topological, whereas the Fubini-Lipatov instanton in the third theory is topologically trivial. The spectral structure in the background of the Fubini-Lipatov instanton can be continuously deformed into that in the flat background, establishing a one-to-one correspondence between the two spectra. However, when considering topologically nontrivial backgrounds for Yang-Mills and $\mathbb{CP}(N\!-\!1)$ theories, the spectrum undergoes restructuring. In these cases, a mismatch between the spectra around the instanton and the trivial vacuum occurs.

\end{abstract}

 \section{Introduction}
 \label{intro}
 
This note is meant to clarify certain confusion regarding calculations of the $\beta$ function by performing the path integral around instantons---or more generally, classical solutions---in theories supporting them. The confusion arises from a specific relationship between the number of zero modes and the asymptotically free contribution to the $\beta$ function. To provide more details, we present a brief introduction.

In the 1950s, Landau and his students~\cite{Landau:1954wla} provided a general explanation for why all field theories known at that time were infrared-free (IRF). The sign of the one-loop graphs, which determine the coupling constant renormalization, is in one-to-one correspondence with the sign of their imaginary parts. This relationship can be demonstrated using the K\"allen-Lehman representation for these graphs. Unitarity implies the positivity of the imaginary parts, which inevitably results in the first coefficients in the $\beta$ functions being {\em positive}, indicating IRF. In four-dimensional theory, IRF was established in arbitrary scalar or Yukawa theories, as well as in Abelian gauge theories with arbitrary matter, bosonic or fermionic.

\begin{figure}[h]
    \centering
     
    \begin{subfigure}[b]{0.3\textwidth}
            
            \centering
            \begin{tikzpicture}[baseline=(a)]

                \begin{feynman}

                    \vertex (a){};
                    \vertex[right=1cm of a] (b);
                    \vertex[right=2cm of b] (c);
                    \vertex[right=1cm of c,dot] (d);

                    \diagram*{
                       (a)--[dotted, very thick](b)--[out=90,in=90,looseness=1.7,gluon, thick](c);
                        (b)--[out=-90,in=-90,looseness=1.7,gluon,thick](c)--[dotted, very thick](d);
                        
                    };
                \end{feynman} 
            \end{tikzpicture}         
            \caption{\label{fig:GhostFreeGaugeA}}
    \end{subfigure}
    \hfil
     \begin{subfigure}[b]{0.3\textwidth}
            
            \centering
            \begin{tikzpicture}[baseline=(b)]

                \begin{feynman}

                    \vertex (a){};
                    \vertex[right=1cm of a] (b);
                    \vertex[right=2cm of b] (c);
                    \vertex[right=1cm of c,dot] (d);

                    \diagram*{
                        (a)--[dotted, very thick](b)--[out=90,in=90,looseness=1.7,dotted, very thick](c);
                        (b)--[out=-90,in=-90,looseness=1.7,gluon,thick](c)--[dotted, very thick](d);
                        
                    };
                \end{feynman} 
            \end{tikzpicture}         
            \caption{\label{fig:GhostFreeGaugeB}}
    \end{subfigure}
      
   \caption{\label{fig:GhostFreeGauge}The dotted lines stand for the Coulomb interaction, the wiggly lines depict transverse gluons.}
     
\end{figure}
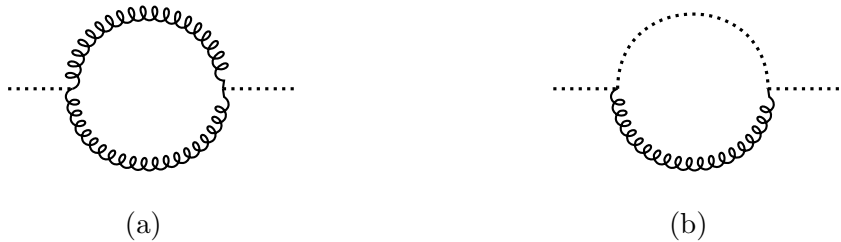

For asymptotic freedom (AF) to occur, the first coefficient of the $\beta$ function must be {\em negative}. The first (and only) theory in four dimensions that has been proven to be asymptotically free is Yang-Mills (YM) theory, which was observed in the late 1960s to early 1970s. The reason for this remarkable phenomenon is the absence of an imaginary part in relevant graphs in unitary gauges, such as the Coulomb gauge, as illustrated in Figure~\ref{fig:GhostFreeGauge}.

Two Feynman graphs for the interaction of (infinitely) heavy quark and antiquark probes were calculated for SU(2) Yang-Mills in \cite{Khriplovich:1969aa}. In Figure~\ref{fig:GhostFreeGaugeA}, a pair of transverse gluons is produced, and this graph has an imaginary part which can be seen by cutting the loop. As in QED, this pair produces screening which leads to IRF. In Figure~\ref{fig:GhostFreeGaugeB}, A similar cut of the loop is impossible since it would go through the Coulomb line, which is, in fact, an instantaneous interaction, leading to the vanishing imaginary part. This graph is responsible for anti-screening, i.e.~AF. The former contribution is 12 times smaller than the latter.

The fact that there are two distinct contributions, one resulting in IRF and the other in AF, is also evident in covariant gauges, such as the background field calculation. If we split the SU(2) gauge field as
\beq
A_\mu^a = {\mc A}_\mu^a +a_\mu^a  \,,
\label{eq:BGFieldMethod}
\eeq
where ${\mc A}_\mu^a$ is the background field and $a_\mu^a$ is the quantum fluctuation, and fix the gauge of $a_{\mu}$ by adding 
to Lagrangian the gauge fixing term of the form,
\be
{\cal L}_{\rm gauge}=-\f{1}{2g_0^2}\,(\mc D^{ab}_\mu a^{b\m})^{2},
\ee
together with the ghost fields $c^{a}$ term,
\be
{\cal L}_{\rm ghost}=-\f{1}{2g_0^2}\,\bar c^a (\mc D_\m \mc D^{\m})^{ab} c^b\,,
\ee
where the covariant derivative $\mc D^{ab}_\mu$ is defined as
\be
\mc D^{ab}_\mu =\delta^{ab}\pt_\mu  +f^{acb} {\mc A}_\mu^c\,,
\ee
then the Lagrangian up to quadratic order in quantum fluctuations---we drop linear terms---takes the form
 \be
 {\cal L}_2 =
 \frac{1}{2g_0^2}\,a^a_\m \l [\eta^{\m \n} (\mc D_\gamma \mc D^{\gamma})^{ab} 
 + 2f^{acb} \mc F^c_{\m \n} \r ] a_\nu^b -\f{1}{2g_0^2}\, \bar c^a  (\mc D_\gamma \mc D^{\gamma})^{ab}  c^b
 \label{eq:QuadraticGluo}.
\ee
Computing the effective action we see that there are different contributions coming from gauge fields and ghosts running in loops
\bea
\label{eq:EffectiveAction}
\mc L _{\rm eff} & = & -\f{1}{4} \mc F_{\mu\nu}^a \mc F^{\mu\nu\, a} \l [ \f{1}{g_0^2} + \f{1}{8\pi^2}\l ( -8+\f{4}{3} \r ) \log \f{M}{\m} - \f{1}{8\pi^2} \f{2}{3} \log \f{M}{\m} \r ] \nn \\
& = &
-\f{1}{4} \mc F_{\mu\nu}^a \mc F^{\mu\nu a} \f{1}{8\pi^2} \l [ \f{8\pi^2}{g_0^2}  - 8 \log \f{M}{\m} + \f{2}{3}  \log \f{M}{\m}  \r ],
\eea
where $M$ is the UV cutoff scale and $\m$ is an arbitrary renormalization scale.
\begin{figure}[h]
    \centering
     
    \begin{subfigure}[c]{0.3\textwidth}
            
            \centering
            \begin{tikzpicture}[baseline=(a)]

                \begin{feynman}

                    \vertex (a);
                    \vertex[right=1cm of a] (b);
                    \vertex[right=1.5cm of b] (c);
                    \vertex[right=1cm of c] (d);

                    \diagram*{
                        (a)--[gluon, thick](b)--[out=90,in=90,looseness=1.7, thick](c)--[gluon, thick](d);
                        (b)--[out=-90,in=-90,looseness=1.7, thick](c)
                    };
                \end{feynman} 
            \end{tikzpicture}         
            \caption{\label{fig:ExternalFieldA}}
    \end{subfigure}
    \hfil
    \begin{subfigure}[c]{0.3\textwidth}
            
            \centering
            \begin{tikzpicture}[baseline=(a)]

                \begin{feynman}

                    \vertex (a);
                    \vertex[above=0.75cm of a] (e);
                    \vertex[below=0.75cm of a] (d);
                    \vertex[below left=1cm of d] (b);
                    \vertex[below right=1cm of d] (c);

                    \diagram*{
                        (b)--[gluon, thick](d)--[gluon, thick](c);
                        (d)--[out=180,in=180,thick,looseness=1.7](e)--[out=0,in=0,thick,looseness=1.7](d)
                    };
                \end{feynman} 
            \end{tikzpicture}         
            \caption{\label{fig:ExternalFieldB}}
    \end{subfigure}
    \hfil
    \begin{subfigure}[c]{0.3\textwidth}
    	\centering
	
            \begin{tikzpicture}[baseline=(a)]

                \begin{feynman}

                    \vertex (a);
                    \vertex[right=1cm of a,dot] (b){};
                    \vertex[right=1.5cm of b,dot] (c){};
                    \vertex[right=1cm of c] (d);

                    \diagram*{
                        (a)--[gluon, thick](b)--[out=90,in=90,looseness=1.6, thick](c)--[gluon, thick](d);
                        (b)--[out=-90,in=-90,looseness=1.6, thick](c)
                    };
                \end{feynman} 
            \end{tikzpicture}
            \caption{\label{fig:ExternalFieldC}}
    \end{subfigure}
   
   \caption{\label{fig:ExternalField}Contributions to the one-loop effective action: Figures~\ref{fig:ExternalFieldA} and \ref{fig:ExternalFieldB}
   represent the ``electric'' interaction of quantum gauge and ghost fields, Figure~\ref{fig:ExternalFieldC} presents magnetic spin interaction of the quantum gauge fields.}
   
\end{figure}
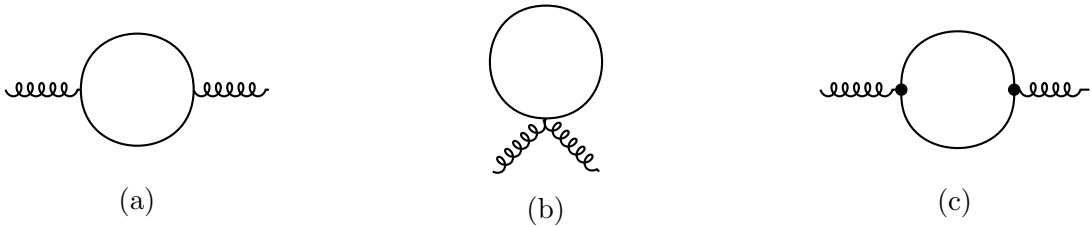
The magnetic spin interaction, represented by the linear term in $\mc F_{\m \n}$ in Equation~\REF{eq:QuadraticGluo} (see Figure~\ref{fig:ExternalFieldC}), produces an AF contribution to the coupling constant renormalization. On the other hand, the ``electric" interaction of the gauge field combined with the contribution from the ghosts (see Figure~\ref{fig:ExternalFieldA} and \ref{fig:ExternalFieldB}), gives rise to an IRF part that is 12 times weaker. Despite the opposite sign of the ghosts' contribution compared to the ``electric" contribution, it should be combined with the latter, canceling the effect of longitudinal modes and preserving only the physical degrees of freedom.

The same phenomenon takes place in two-dimensional sigma model, say, $\mathbb{CP}(1)$, see Appendix~\ref{app:CP1Model} for details. 
It is known that the model is asymptotically free. However, as we will show below, if we take into account only the Feynman graph which does have imaginary part, the result will be IR free. In much the same way as in 4D Yang-Mills, to make the first coefficient of the $\beta$ function negative (to ensure asymptotic freedom) we need to add something else. In $\mathbb{CP}(1)$  this``something else" is the tadpole graph with no imaginary part.

Thus, in both cases mentioned above the ``Landau theorem" is still valid: if we include in perturbation theory only the diagrams for polarization operator with non-vanishing imaginary parts, the $\beta$ function turns out IR free. It is the diagrams with no imaginary parts which convert IRF into AF. For brevity, we will refer to the latter contributions as {\em extra.}

There exist an alternative method of calculation of the $\beta$ function -- using instanton background instead of flat. Technically, it is more cumbersome but also more instructive. For the BPST instanton this calculation was carried out by 't Hooft in great detail. Needless to say, both methods give one and the same $\beta$ function. The instanton calculation shows that it is the zero mode contribution which is responsible for 
for asymptotic freedom. Exactly the same happens in the $\mathbb{CP}(1)$ instanton analysis.

It is worth recalling 't Hooft's instanton measure calculation~\cite{tHooft:1976snw}. 
For the SU(2) gauge field, the measure takes the form
\beq
d\mu_{\rm inst}  ={\rm const}\times 
\, \int \frac{d^4x_0\, d\rho}{\rho^5} \big(M\rho \big)^{8} \left( \frac{8\pi^2}{g^2_0}\right)^{4}
\, \exp\left(- \frac{8\pi^2}{g^2_0}  + \Delta_{\rm gl} + \Delta_{\rm gh}
\right),
 \label{3}
\eeq
where the exponent ${8\pi^2}/{g^2_0}$ is the action of  the classical solution (the instanton), $g_0^2$ is the ``bare" coupling constant at the UV scale $M$, which is the Pauli-Villars regulator mass, and $\rho$ is the instanton size. All pre-exponential factors in (\ref{3}) come from  the zero modes. Furthermore, 
$\Delta_{\rm gl}+ \Delta_{\rm gh} $ in the exponent represent the {\em bona fide} quantum corrections in the instanton background, which take into account only non-zero modes
\beq
\Delta_{\rm gl}+ \Delta_{\rm gh} = -\frac 23 \log M\rho \, .
\label{4}
\eeq
The zero modes emerge due to the spin term in (\ref{eq:QuadraticGluo}). Combining (\ref{3}) and (\ref{4}), we conclude that at one loop the  renormalized coupling is given by
\beq
 \frac{8\pi^2}{g^2(\rho)}=  \frac{8\pi^2}{g^2_0}- 8\log M\rho +\frac 23 \log M\rho,
\eeq 
which, corresponds to the AF and coincides with \REF{eq:EffectiveAction}, as expected.

Here comes an interesting peculiarity of the last computation. We observe that the anti-screening contribution, coming from graphs without an imaginary part (Figure~\ref{fig:GhostFreeGaugeB}), is given exactly by the zero modes in the instanton computation.\footnote{\,This fact is especially remarkable for supersymmetric theories~\cite{Novikov:1983uc}, where non-zero modes' contribution is absent, therefore, the zero modes provide the full result.} The determinant of non-zero modes, on the other hand, coincides with the ``normal" screening contribution given by the graphs in Figure~\ref{fig:GhostFreeGaugeA}, which can be easily reconstructed from the corresponding imaginary part (that is in turn positive). It was also observed, although it was left without an explanation, in~\cite{Vainshtein:1981wh}, that the contribution of non-zero modes can be computed via a small field expansion~\REF{eq:BGFieldMethod} if one disregards the spin-dependent diagrams, Figure~\ref{fig:ExternalFieldC}.

A technical reason for the phenomenon occurring in four-dimensional YM theory is provided by 't Hooft in~\cite{tHooft:1976snw}. He showed that the non-zero eigenvalues for fluctuations around the instanton do not depend on the spin. Therefore, computing the product of non-zero modes is equivalent to computing the determinant for a $4-2=2$ component scalar (remember, we are considering pure YM theory without fermions). 

The above two examples refer to topologically non-trivial classical solutions. A non-topological instanton (more precisely, bounce)
exists in the Symanzik version of the $\lambda\phi^4$ theory (to be referred to as S$\lambda\phi^4$). The Symanzik version differs from the {\em bona fide} $\lambda\phi^4$ by an artificial 
change of the sign of the coupling constant $\lambda$ in the Lagrangian. Thus, S$\lambda\phi^4$ has no ground state. 

However, in the Euclidean space it does have the Fubini-Lipatov instanton. The non-physical nature of S$\lambda\phi^4$ is irrelevant in perturbation theory near $\phi=0$. The one-loop graph
for $\lambda$ renormalization is unique and has one and the same positive imaginary part, both in $\lambda\phi^4$  and S$\lambda\phi^4$.   However, artificially changing the sign of the tree $\lambda\phi^4$ vertex we reverse the running law of the renormalized coupling, from Landau's IR freedom   to  AF. This is a ``fake" asymptotic freedom, though. 

Comparison of the calculations of the $\beta$ function in the flat and FL instanton backgrounds reveals this fact unequivocally -- there are no extra contributions and FL instanton zero modes play no special role. The conceptual difference is due to the fact that in S$\lambda\phi^4$ one can see that the spectral flow is continuous in passing from the flat background to that of the FL instanton.

So far, the calculation of the $\beta$ function in the FL instanton background was not carried out. Although it was analyzed  many times, in works~\cite{Komarova:2005qe, Garbrecht:2018rqx} and others the $\lambda$ renormalization in these analyses is just taken from the perturbation theory over the flat vacuum. The calculation we perform below is  the first one performed in the FL instanton background. 

Our study was initiated by observing the difference in the relationship between the zero modes and beta functions in the case of the Fubini-Lipatov (FL) instanton compared to the BPST and Polyakov-Belavin $\mathbb{CP}(1)$ instantons. We discovered that this difference arises due to the non-topological nature of the FL instanton.

Before concluding the introduction let us make two remarks. The most transparent way to trace the spectral flow is through introduction of massless fermions. The external anomaly in the divergence 
of the axial current
serves as a detector of level crossing. The chiral anomaly is present in Yang-Mills theory and in $\mathbb{CP}(1)$; on the other hand
there is no analog of this phenomenon in $\lambda\phi^4$. Moreover, if we supersymmetrize the theories under consideration, all non-zero modes 
cancel each other in the instanton background -- zero modes fully determine the $\beta$ function.\footnote{\,By introducing Dirac fermions and observing the presence of an external chiral anomaly in the given background, we can determine the index of the Dirac operator. If the index does not vanish, it indicates the appearance of extra fermion spectral modes in the topologically {\em nontrivial} background. Upon supersymmetrization and given the background field preserving at least a part of the supercharges, the extra fermion zero modes will be accompanied by extra boson zero modes. In logarithmically renormalized theories, the boson zero modes will exhibit asymptotic freedom (AF), while the fermion modes will contribute to infrared freedom (Landau zero charge).} This is explicit in Yang-Mills and $\mathbb{CP}(1)$.
It is important to realize that S$\lambda\phi^4$ cannot be supersymmetrized (as opposed to ``normal"  $\lambda\phi^4$ which can be supersymmetrized but has no instantons.)

Our second remark concerns three-dimensional Yang-Mills theory in Euclidean space (equivalent to static Georgi-Glashow model).
As was noted by Polyakov, it has instantons usually referred to as monopole instantons in this context.\footnote{\,In fact, the monopole instantons  lead to the Polyakov confinement in 3D.} They are topologically non-trivial solutions and do have zero modes.
However, the theory is super-renormalizable and regularized in IR by the vacuum expectation value of the scalar field.\footnote{\,Note that chiral anomalies are only present in two and four dimensions, as there is no chirality in three dimensions.} There is no {\em bona fide} running in the UV too.

The organization of the paper is as follows. In Section~\ref{setu}, we address the $\lambda\phi^4$ theory and explain how the Fubini-Lipatov (FL) instantons contribute to the analysis of the $\beta$ function. To do this, we compare the instanton computations in this theory with well-known examples such as the BPST instanton in YM theories (see References~\cite{Novikov:1983uc, Vainshtein:1981wh}) and the Polyakov-Belavin instanton~\cite{Polyakov:1975yp} in the two-dimensional 
$\mathbb{CP}(1)$ model (see Appendix \ref{app:CP1Model}). A key feature of the latter spectra is the emergence of zero modes in the instanton background that are treated as ``extra" in the sense explained above.  The necessary condition for the generation of  extra modes is the nontrivial topology of the instantons under consideration. The FL instanton is, on contrary, non-topological. We demonstrate that in this case there are no extra emerging modes. The spectrum in the flat vacuum is continuously deformable  into that in the FL instanton background. There are no non-dispersive graphs in S$\lambda\phi^4$.

To this end we discretize the spectrum of the eigenmodes in the FL background by putting the theory on a sphere, which is the simplest and most transparent way of discretization for the problem at hand. This is discussed in Section~\ref{sec:Sphere-PV}. Our conclusions are summarized in Section~\ref{conclu}. Appendix~\ref{app:CP1Model} highlights the similarities between the instanton calculations of the $\beta$ functions in four-dimensional YM theory and the two-dimensional sigma model. Appendices \ref{app:HardCutoff} and \ref{app:CutoffSphere} provide some technical details.

\section{Setup}
\label{setu}

We are studying an $O (N)$ invariant Euclidean scalar field theory in $d=4$ dimensions
\be
\label{eq:LagrangianPhi4}
\mc L = \f{1}{2} (\p \phi_a)^2 - \f{g_0}{4!} (\phi^2_a)^{2}, ~~a=1,\dots, N, ~~ g_{0}>0.
\ee
Despite the negative potential, this theory is well-defined perturbatively, meaning that there are no instabilities arising at perturbative level. At the same time such a non-standard choice of the coupling leads to an interesting feature of the theory, it is asymptotically free~\cite{Symanzik}.
There are several ways to compute the $\b$-function, for instance, using Feynman diagrams. For our purposes, though, it is more instructive to show explicitly, how to find the one-loop $\b$-function by computing the effective potential.

Expanding fields around a non-trivial profile
\be
\phi_N = \phi_0 + \vp_N, ~~ \phi_a = \vp_a, ~~ a \neq N,
\ee
leads to (we tacitly assume the presence of necessary sources)
\be
\label{eq:quadraticAction}
S= S_0 + \int d^4 x \l [\f{1}{2} (\p \vp_N)^2 + \f{1}{2} \,(\p \vp_a)^2 
- \f{g_0}{4} \,\phi_0^2 \vp_N^2 - \f{g_0}{12}\,\phi_0^2 \vp_a^2 \r], ~~ a \neq N.
\ee
As a result the effective action becomes
\be
\label{eq:EffectiveActionBare}
\G[\phi_0] = S_0 + \f{1}{2} \,\Tr \log\l [ -\p^2 - \f{g_0 \phi_0^2}{2}\r ] + \f{N-1}{2}\,\Tr \log\l [ -\p^2 - \f{g_0 \phi_0^2}{6}\r ].
\ee
Computing determinants is tantamount to summing all one-loop graphs in external field.
\begin{figure}[h]
    \centering
     
    \begin{subfigure}[c]{0.3\textwidth}
            
         \centering

	\begin{tikzpicture}
		\draw[black, very thick] (0,0) circle (1){};
		\end{tikzpicture}

            \caption{\label{fig:EffectiveActionPhiA}}
    \end{subfigure}
    \hfil
    \begin{subfigure}[c]{0.3\textwidth}
            
            \centering
           \begin{tikzpicture}
		\draw[black, very thick] (0,0) circle (1){};
		\filldraw[white, thick] (-1,0) circle (2pt){};
		\draw[black, thick] (-1,0) circle (2pt){};
		\draw[decorate,decoration=crosses,thick] (-1,0pt) - - (-1,1pt);
	\end{tikzpicture}        
            \caption{\label{fig:EffectiveActionPhiB}}
    \end{subfigure}
    \hfil
    \begin{subfigure}[c]{0.3\textwidth}
    	\centering
             \begin{tikzpicture}
		\draw[black, very thick] (0,0) circle (1){};
		\filldraw[white, thick] (1,0) circle (2pt){};
		\draw[black, thick] (1,0) circle (2pt){};
		\draw[decorate,decoration=crosses,thick] (1,0pt) - - (1,1pt);
		\filldraw[white, thick] (-1,0) circle (2pt){};
		\draw[black, thick] (-1,0) circle (2pt){};
		\draw[decorate,decoration=crosses,thick] (-1,0pt) - - (-1,1pt);
	\end{tikzpicture} 
            \caption{\label{fig:EffectiveActionPhiC}}
    \end{subfigure}
   
   \caption{\label{fig:EffectiveActionPhi}}   
\end{figure}

Only the first three graphs in Figure~\ref{fig:EffectiveActionPhi} are divergent. When considering a constant background $\phi_0=\phi_c$, the computation becomes straightforward. However, it is necessary to regularize determinants. We introduce Pauli-Villars regulators $\Phi_{a,i}$---three regulators for each $\vp_{a}$ (including $a=N$), with statistics $c_{a,i}$ and masses $M_{a,i}$ (where $i=1,2,3$). We choose the same masses and statistics for any index $a$ and denote them as $M_{a,i} = M_i$ and $c_{a,i} = c_i$. Consequently, the regulated effective action is given by
\bea
\G_{R}[\phi_c] &\!\! \!\!=\!\!\!\! &-\!\int\!\! d^4 x\, \f{g_0\phi_c^4}{4!} -\f{1}{2} \sum_i c_{i} \Tr \log\l ( -\p^2 - \f{g_0 \phi_c^2}{2} + M_{i}^2\r ) + \f{1}{2} \Tr \log\l ( -\p^2 - \f{g_0 \phi_c^2}{2}\r ) \nn \\
&& \hspace{0.01cm} - \f{N-1}{2} \sum_i c_{i} \Tr \log\l ( -\p^2 - \f{g_0 \phi_c^2}{6} + M_{i}^2\r ) + \f{N-1}{2} \Tr \log\l ( -\p^2 - \f{g_0 \phi_c^2}{6} \r ).
\eea
Using the fact that that 
\bea
\int_{\mu}^ \Lambda \f{d^4 k}{(2\pi)^4} \log \l ( k^2 +m^2 \r )&\!\! =\!\!&\frac{1}{32\pi^{2}}\bigg\{\!\Lambda^{4}\bigg(\log \l ( \Lambda^2  +m^2 \r ) 
-\frac{1}{2}\bigg)-\mu^{4}\bigg(\!\log \l ( \mu^2 +m^2 \r )-\frac{1}{2}\bigg) \nn\\
&&\hspace{1.2cm}+m^{2}\l ( \Lambda^2 -\mu^2 \r ) -m^{4}\log\frac{ \Lambda^2  +m^2 }{\mu^{2}+m^{2}}\bigg\}
\eea
and taking the $\Lambda=\infty$ limit we arrive at (for $\mu\gg g\phi_{c}^{2}$ and omitting finite terms) 
\be
\label{eq:constant}
\G_{R}[\phi_c] = \int d^4 x \l [ -\f{g_0\phi_c^4}{4!} \l ( 1 + \f{g_0}{32\pi^2} \f{N+8}{3}\log \f{M}{\m} \r ) + b_4 M^4 +  b_2 M^2 g\phi_c^2 \r ],
\ee
where $b_{4}$ and $b_{2}$ are numerical constant and we used that
\be
\sum c_i =1, ~~ \sum c_i M_i ^2 =0, ~~ \sum c_i M_i^4 =0, ~~ \log M \stackrel{\rm def}{=} \sum c_i \log M_i\,.
\ee

The terms proportional to $M^4$ and $M^2$ are absorbed in the cosmological constant and mass counter terms. The $\log M$ dependence means that the renormalized at a scale $\m$ coupling is given by
\be
\label{eq:trivialBackground}
g(\m) = g_0 \l ( 1 + \f{g_0}{16\pi^2} \f{N+8}{3} \log \f{M}{\m} \r ),
\ee
which corresponds to asymptotically free behavior
\be
\label{eq:betaFunction}
\b_{g}(g) = -\f{g^2}{16\pi^2}\f{N+8}{3}\,.
\ee

\paragraph{Fubini-Lipatov instanton} There is a classical solution~\cite{Fubini:1976jm,Lipatov:1976ny} in this theory
\be
\phi_N =\phi_{FL} \equiv 4 \, \sqrt{\f{3}{g_0}} \, \f{\rho}{r^2+\rho^2}, ~~ \phi_i=0, ~~i\neq N,
\ee
which is called the Fubini-Lipatov instanton (in fact it is a bounce). Similarly to Yang-Mills theory, parameter $\rho$ is the size of the instanton, which is not fixed classically. When computing the path integral around this configuration, that we schematically denote as
\be
I_{FL} = \int \mc D \vp e^{-S[\phi_{FL}+\vp]},
\ee
one should be careful, for there are several zero modes. Namely, there are five zero modes corresponding to the broken space-time symmetries. Classically, the Lagrangian \REF{eq:LagrangianPhi4} is conformally invariant, which means that the instanton not only breaks translations but also dilations. Additionally, there are zero modes associated with the breaking of the internal symmetry $SO(N) \to SO (N-1)$, with a total of $N-1$ such modes. Furthermore, there is one negative mode with eigenvalue $\lambda_-$. By factoring out these modes, as usual replacing the zero modes with integration over moduli $dX$, the regulated integral becomes

\be
\label{eq:PV}
I _{FL}= \int \sqrt{ \f{ \prod_i (M^2_{i} +\lambda_-)^{c_{i}} }{ \lambda_-}} \, dX \, \prod_i M_{i}^{(N+4)c_{i}} 
 \, \exp \l \{-\G_R[\phi_{FL}] \r \},
\ee
with
\bea
\label{eq:GammaRegF}
\G_R[\phi_{FL}]& = &\int d^4 x \, \l ( \f{1}{2} (\p \phi_{FL})^2 - \f{g_0\phi_{FL}^4}{4!} \r ) \nn \\ 
&& - \f{1}{2} \sum_i  c_{i} \Tr \log\l ( -\p^2 - \f{g_0 \phi_{FL}^2}{2} + M_{i}^2\r ) + \f{1}{2} \Tr \log\l ( -\p^2 - \f{g_0 \phi_{FL}^2}{2}\r ) \\
&& - \f{N-1}{2} \sum_i c_{i} \Tr \log\l ( -\p^2 - \f{g_0 \phi_{FL}^2}{6} + M_{i}^2\r ) + \f{N-1}{2} \Tr \log\l ( -\p^2 - \f{g_0 \phi_{FL}^2}{6} \r ) . \nn
\eea
This expression illustrates the point we discussed in the Introduction. If we were to assume that the leading $\log M$ behavior of \REF{eq:GammaRegF} could be obtained by neglecting the spatial dependence of $\phi_{FL}(x)$, we would arrive at an expression identical to~\REF{eq:constant}. However, there is an additional contribution to the $\log M$ dependence arising from the zero and negative modes in \REF{eq:PV}. Therefore, this naive treatment would fail to yield the correct result in \REF{eq:betaFunction}. Consequently, we need to exercise more diligence when computing determinants around the instanton background. In the next section, we discretize the spectrum by putting the theory on a sphere\footnote{\,This approach is similar to what is done in~\cite{tHooft:1976snw} for Yang-Mills theory and in~\cite{Jevicki:1977yd} for the $\mathbb{CP}(1)$ sigma model. Mathematically, it means that we appropriately choose the measure with respect to which eigenfunctions are orthogonal.}. This will enable us to accurately perform the necessary computations.

\section{Mapping on a sphere \label{sec:Sphere-PV}}

We find it beneficial to temporarily keep the number of dimensions $d$ general. Our theory can be obtained from the following (Euclidean) Lagrangian on $\mathbb R^d$
\be
\label{eq:FlatLagrangian}
\mc L = \f{1}{2} (\p \phi_a)^2 - \f{g_0}{\G(5+2\a)} (\phi^2_a)^{2+\a(d)}, ~~ \a(d) = \f{4-d}{d-2}, ~~ g_0>0.
\ee
The Fubini-Lipatov instanton in this case can be written as
\be
\label{eq:Fubini_plane}
\phi_N =\phi_{FL} \equiv  \l [ \f{d(d-2)}{g_0}\G\l ( \f{2d}{d-2} \r ) \r ]^\f{d-2}{4} \l( \f{\rho}{r^2+\rho^2} \r )^{\f{d}{2}-1}, ~~ \phi_a=0, ~~a\neq N.
\ee
For each single choice of $\rho$, using the power of conformal invariance, we put the theory \REF{eq:FlatLagrangian} on a sphere $\mathbb S^d$ by employing the standard stereographic projection
\be
\sqrt{x_1^2+\dots + x_d^2} = \rho\cot \f{\t_d}{2}.
\ee
To simplify notations we will set $\rho=1$, which can be easily restored simply using dimensional analysis. 
As a result the theory becomes
\be
S_{d} = \int d \Omega_d \l [ \f{1}{2} \l ( \nabla \phi_a \r )^2 + \f{d(d-2)}{8} \phi_a^2 - \f{g}{\G(5+2\a)} (\phi^2_a)^{2+\a(d)} \r ],
\ee
where integration runs over $d$-dimensional sphere with volume
\be
\Omega_d = \f{2\pi^{(d+1)/2}}{\G\l ( \f{d+1}{2}\r )}\,.
\ee
In this case the instanton is just the maximum of the potential
\be
\label{eq:Fubini_sphere}
\phi_N =\phi^s_{FL} \equiv  \l [ \f{d(d-2)}{4g}\G\l ( \f{2d}{d-2} \r ) \r ]^\f{d-2}{4},~~ \phi_a=0,  ~~~~a\neq N
\ee
which is clearly in agreement with \REF{eq:Fubini_plane}.

It is beneficial to first compute the effective action for an arbitrary constant profile. Expanding the action we get at quadratic order
\bea
S_d[\phi_c+\vp]& = & \Omega_d \l [  \f{d(d-2)}{8} \phi_c^2 - \f{g}{\G(5+2\a)} \phi_c^{4+2\a(d)} \r ] \\
&&+ 
\int d \Omega_d \l [  \f{1}{2} \l ( \nabla \vp_N \r )^2+\f{1}{2} \l ( \nabla \vp_a \r )^2 + \f{\overline m^2(d)}{2} \vp_N^2
+\f{m^2(d)}{2} \vp_a^2 \r ], \nn
\eea
where $a\neq N$ and $\phi_c$- and $d$-dependent masses for fields are
\bea
\overline m^2_d(\phi) & = & \f{d(d-2)}{4}-\f{g \phi^{\f{4}{d-2}}}{\G\l ( \f{d+2}{d-2} \r )} \\
m^2_d (\phi) & = & \f{d(d-2)}{4}-\f{g \phi^{\f{4}{d-2}}}{\G\l ( \f{2d}{d-2} \r )}.
\eea
Using that eigenvalues of the Laplacian on a $d$-dimensional sphere are give by 
\be
J_\ell(d) = \ell (\ell+d-1),
\ee
we find the spectrum
\bea
\label{eq:SphereSpectrum}
\overline \lambda_\ell (d) & = & J_\ell(d)+\overline m^2_d(\phi_c), \\
\lambda_\ell (d) & = & J_\ell(d)+m^2_d(\phi_c),
\eea
with multiplicity
\be
\n_\ell (d) = \f{(2\ell+d-1)(\ell +d-2)!}{\ell! (d-1)!}.
\ee
Pauli-Villars regularization leads to the following expression for the effective action
\bea
\label{eq:Gamma1LoopPV}
\G_d[\phi_c] & = & S_d[\phi_c] + \f{1}{2} \sum^\infty_{\ell=0} \n_\ell(d) \l [ \log \bar \lambda_{\ell} (d) + (N-1) \log \lambda_{\ell} (d)\r ] \\
&& - \f{1}{2} \sum_i c_i \sum^\infty_{\ell=0} \n_\ell(d) \l [ \log \l ( \bar \lambda_{\ell} (d) + M_i^2 \r ) + (N-1) \log \l ( \lambda_{\ell} (d) + M_i^2 \r ) \r ]. \nn
\eea
By defining
\be
\label{eq:SigmaDef}
\s_d(a,\ell_0) = \f{1}{2} \sum^\infty_{\ell=\ell_0} \n_\ell(d) \l \{ \log \l [ J_{\ell} (d) + a \r ] - \sum_i c_i \log  \l [ J_{\ell} (d) + a + M_i^2 \r ]\r \},
\ee
we can rewrite the effective action as (neglecting finite terms)
\be
\label{eq:GammaL0}
\Gamma_d [\phi_c] = S_d[\phi_c] + \s_d(\bar m^2_d,\ell_0)+ (N-1) \s_d(m^2_d,\ell_0)  - N \log M \, \sum_{\ell=0}^{\ell_0-1} \n_\ell (d).
\ee
Clearly, \REF{eq:GammaL0} should not depend on $\ell_0$, because it was introduced artificially by splitting the sum of eigenvalues into the bottom $\ell_0$ modes and the rest. This split is meant for illustrating that the final result does not depend on whether we treat the zero modes separately or not. 

The divergent part of \REF{eq:SigmaDef}, which we are interested in, can be computed using the Euler-Maclaurin summation formula, giving
\be
\s_4 (a,\ell_0) = b_4M^4+ b_2M^2-\l [ \f{a(a-4)}{12} +\f{29}{90} \r] \log M + \f{1}{12} \ell_0(\ell_0+1)^2(\ell_0+2)\log M.
\ee
Using that 
\be
\sum_{\ell=0}^{\ell_0-1} \n_\ell (d) = \f{(2\ell+d-2)(\ell+d-2)!}{d! (\ell-1)!},
\ee
we see that, indeed, $\ell_0$ dependence of the effective action disappears and we obtain
\be
\label{eq:Gamma4Full}
\Gamma_4 [\phi_c] = \f{8\pi^2}{3}\phi_c^2-\f{g_0 \pi^2}{9}\phi_c^4 \l ( 1+ \f{N+8}{3}\f{g_0}{16\pi^2} \log M \r ) + \f{N}{90} \log M + b_4M^4+ b_2M^2.
\ee
We conclude that, as before, the renormalized coupling is given by \REF{eq:trivialBackground}. Therefore, the beta function is the same as in \REF{eq:betaFunction}. Other terms can be absorbed into renormalization of operators involving the curvature and cosmological constant.

Computing the path integral around the instanton configuration in this case is not technically different from computing the effective action for a constant profile. The only complication, compared to $\mathbb R^d$, arises from the presence of other operators contributing to the $\log M$ dependence, as we can see from \REF{eq:Gamma4Full}. Evaluating the effective action on the instanton background \REF{eq:Fubini_sphere}, modulo polynomial terms, we have
\be
\Gamma_4 [\phi_{FL}^s] = \f{16 \pi^2}{g_0^2} - \f{N+8}{3} \log M + \f{N}{90} \log M
\ee
It is evident that the coefficient in front of $\log M$ does not correspond to the coupling renormalization. To circumvent this minor issue, we observe that the problematic term in \REF{eq:Gamma4Full} is not field dependent. Therefore, normalizing the determinant to that of the trivial background would resolve the problem. Physically, this procedure corresponds to calculating the relative free energy, similar to what was done in~\cite{tHooft:1976snw}. Simple computation reveals
\be
\label{eq:EA4DiffLambda}
\G_4[\phi_{FL}^s] - \G_4[0] = \f{16\pi^2}{g_0^2} - \f{N+8}{3} \log M,
\ee
which can now be used to find the beta function.

\section{Discussion and conclusion \label{conclu}}

To analyze the result let us take a closer look at the structure of our computation. The fluctuations around the Fubini-Lipatov instanton for $N=1$ are encapsulated in the following pre-exponential factor
\be
\mc P_{FL}= \int dX\, M^{6}
\l [ \f{\det' \l ( \mc M_\mc I+M^2\r )}{\det' \mc M_\mc I} \, \f{\det \mc M_0}{\det  \l ( \mc M_0+ M^2\r ) } \r ] ^{1/2},
\label{eq:FLPreexp}
\ee
where $\mc M _0$ and $\mc M_\mc I$ are the corresponding differential operators around the trivial profile and the Fubini-Lipatov instanton, respectively. It is important to note that this expression is only schematic, for it neglects the presence of several regulator fields. Nevertheless, it serves to illustrate what is going on.

As we can see, there are two sources of the UV cutoff ($M$) dependence of the pre-exponential factor. The first contribution, $M^6$, arises from the zero (and the negative) modes, while the second contribution comes from the ratio of products involving only the positive modes.
\be
\l [ \f{\det' \l ( \mc M_\mc I+M^2\r )}{\det  \l ( \mc M_0+ M^2\r ) } \r ] ^{1/2}.
\ee
As a result the relevant part of \REF{eq:FLPreexp} is given by
\be
\label{eq:RatioRegulatorsM}
\mc R_{FL} = M^6 \l [ \f{\det' \l ( \mc M_\mc I+M^2\r )}{\det  \l ( \mc M_0+ M^2\r ) } \r ] ^{1/2}.
\ee
The ratio of determinants should be understood as the ratio of products with a common cutoff. Namely,
\be
\f{\det' \l ( \mc M_\mc I+M^2\r )}{\det  \l ( \mc M_0+ M^2\r ) } = 
\f{\prod_{\ell=2}^\Lambda \l ( \lambda_\ell^{\mc I}+M^2\r ) ^{\n_\ell/2}}
{\prod_{\ell=0}^\Lambda\l ( \lambda_\ell^{0}+M^2\r ) ^{\n_\ell/2}}.
\ee
The mismatch in the number of modes in the numerator and denominator (products starting from $\ell=2$ and $\ell=0$ correspondingly) is precisely compensated by the $M^6$ factor. Indeed, with one $\ell=0$ mode and five $\ell=1$ modes we have
\be
{\prod_{\ell=0}^1\l ( \lambda_\ell^{0}+M^2\r ) ^{\n_\ell/2}}=(\lambda_{0}^{0}+M^{2})^{1/2}(\lambda_{1}^{0}+M^{2})^{5/2}\approx M^{6}\,.
\ee
Hence, with $\log M$ precision we get
\be
\label{eq:FLSchematicNonZero}
\mc R_{FL} = \f{\prod_{\ell=0}^\Lambda \l ( \lambda_\ell^{\mc I}+M^2\r ) ^{\n_\ell/2}}
{\prod_{\ell=0}^\Lambda\l ( \lambda_\ell^{0}+M^2\r ) ^{\n_\ell/2}}.
\ee
The above consideration demonstrates that the spectral flow, when moving from a trivial background to the instantonic one, supports the continuity of levels. No new levels appear; instead, a few low levels from the trivial background shift downwards to become zero or negative modes. However, this is not always the case for all theories. In the case of YM theory and the non-linear $\mathbb{CP}(1)$ model, the zero modes around the instanton are genuinely ``new", meaning they do not have counterparts in the spectrum around the trivial background. As a result, there is no one-to-one correspondence between the modes as in equation \REF{eq:FLSchematicNonZero}.

When all modes are paired as in Equation \REF{eq:FLSchematicNonZero}, the $M$-dependence can be easily determined by analyzing the asymptotic behavior of the products.  In particular, if we are interested in the $\log M$ contribution, it is sufficient to identify the terms proportional to $\ell^{-1}$ in the summand (see Appendix~\ref{app:CutoffSphere})
\be
\f{1}{2}\,\n _\ell \log \lambda^{\mc I, 0}_\ell.
\ee

Let us return to three-dimensional Higgsed YM (more exactly, the Georgi-Glashow model), where, as was mentioned in Introduction, one can consider instanton-monopole as a background field. This three-dimensional example has no infrared problem since all physical fields (except ``photon") acquire masses. Since it is super-renormalizable, it has no scale anomaly. If we added fermions there would be no chiral anomaly too since chirality is not defined in three dimensions. We expect that studying the structure of the level flow we will see that the spectrum in this model is similar to the one around the FL instanton: no extra levels appear in passing from the flat vacuum to the instanton background. An indirect indication of the essential difference between the spectra is the fact that the BPST instanton in $4D$ and the Polyakov-Belavin instanton in $2D$ can be supersymmetrized leading to complete cancelation of all non-zero modes. The FL instanton cannot be embedded in any supersymmetric model.

The final comment concerns performing computations directly in flat space. There is no conceptual obstacle to evaluating Eq.\,\REF{eq:RatioRegulatorsM} without mapping the theory onto the sphere. However, it should be noted that, in general, this computation does not simplify to merely computing the effective potential around a constant background and then evaluating it on the instanton. The coordinate dependence of the instanton background necessitates computing the effective action, rather than just the effective potential. In four dimensions, due to a peculiar feature---at one loop the field $\phi$ is not renormalized---only one operator ($\phi^4$) in the Lagrangian receives $\log M$ corrections. Hence, the full $\log M$ dependence can be extracted from the effective potential. However, it is easy to see that evaluating only the effective potential in six dimensions leads to incorrect results (see Appendix~\ref{app:CutoffSphere}).

\section{Acknowledgements}

We are very much thankful to Tony Gherghetta who essentially initiated this study asking the proper questions.

This work is supported in part by DOE grant DE-SC0011842. AM, whose work was partially supported by the Simons Foundation Targeted Grant for the Fine Theoretical Physics Institute, acknowledges the warm hospitality extended by the W. Fine Theoretical Physics Institute at the University of Minnesota. AV is grateful  for the hospitality to the Kavli Institute for Theoretical Physics at the University of California, Santa Barbara where his work was supported by the National Science Foundation under Grant No. NSF PHY-1748958.

\appendix

\section{\boldmath{$\mathbb{CP}(1)$} sigma model \label{app:CP1Model}}

In case of $\mathbb{CP}(N\!-\!1)$ the $N$-dependence of the one-loop beta function is just a simple 
proportionality to $N$. So consideration of $\mathbb{CP}(1)$ sigma model ($N=2$) is sufficient for our purposes.

The model in question is given by the following Lagrangian in two-dimensional Euclidean space
\be
\mc L _{CP} = \f{2}{g_0^2} \,\f{\p_\m \bar\phi \p_\m \phi}{(1+\bar\phi \phi)^2}.
\label{eq:lagrCP}
\ee
Introducing complex coordinates
\be
z= x_0 + i x_1, ~~ \bar z= x_0 - i x_1,
\ee
it is straightforward to find the instanton solution
\be
\phi = \f{a}{z-b},
\ee
with $a$ and $b$ complex parameters. Thus, there are four zero modes. As a result the instanton measure is given by
\beq
dX_{CP}  ={\rm const}\times 
\, \int d^2a \, d^2 b \, M^4 \left( \frac{4\pi}{g^2_0}\right)^{2}
\, \exp\left(- \frac{4\pi}{g^2_0}  + \Delta' \right),
\eeq
with $\D'$ representing the contribution of non-zero modes, see~\cite{Jevicki:1977yd},
\be
\D' = -2 \log M.
\ee
Collecting all $\log M$ contributions we get that the renormalized coupling becomes
\be
\f{4\pi}{g^2}=\f{4\pi}{g_0^2}-4 \log M +2 \log M.
\label{eq:CP1g2Renormalization}
\ee

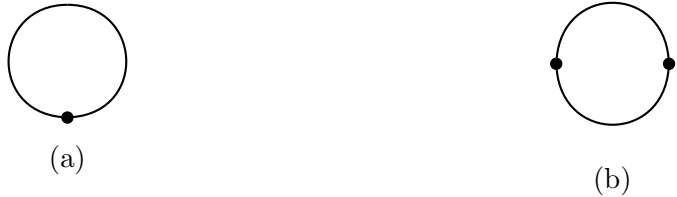
\begin{figure}[h]
    \centering
     
    \begin{subfigure}[c]{0.3\textwidth}
            
            \centering
            \begin{tikzpicture}[baseline=(a)]
		
		\begin{feynman}

                    \vertex [dot](a){};
                    \vertex[above=1.5cm of a] (d);

                    \diagram*{
                        (a)--[out=175,in=180,thick,looseness=1.7](d)--[out=0,in=5,thick,looseness=1.7](a)
                    };
                \end{feynman}

            \end{tikzpicture}         
            \caption{\label{fig:CP1DiagramsA}}
    \end{subfigure}
    \hfil
    \begin{subfigure}[c]{0.3\textwidth}
            
            \centering
            \begin{tikzpicture}[baseline=(a)]

                \begin{feynman}

                    \vertex [dot](a){};
                    \vertex[right=1.5cm of a,dot] (b){};

                    \diagram*{
                        (a)--[out=85,in=95,looseness=1.7, thick](b)--[out=-95,in=-85,looseness=1.7, thick](a);
                    };
                \end{feynman}
              
            \end{tikzpicture}         
            \caption{\label{fig:CP1DiagramsB}}
    \end{subfigure}
   
   \caption{\label{fig:CP1Diagrams} One-loop diagrams in the effective Lagrangian: (a) tadpole and (b) dispersive. The bold blobs refer to the full $\mathbb{CP}(1)$ Lagrangian, external lines are not shown. }
\end{figure}

We would like to compare this with 
the usual perturbative computation around the trivial background. 
To this end we turn to the one-loop calculations given in the Appendix C of Ref.~\cite{Novikov:1984ac}.
Referring for details to \cite{Novikov:1984ac} let us note that there are two pieces of one-loop addition 
the original Lagrangian \REF{eq:lagrCP}, see diagrams in Fig.~\ref{fig:CP1Diagrams}.

The first one produced by just a tadpole, Fig.~\ref{fig:CP1Diagrams}a, has the form
\be
\Delta \mc L_{\rm tadpole}=\p_\m \bar\phi \p_\m \phi\bigg[\frac{6\bar\phi \phi}{(1+\bar\phi \phi)^{2}} -\frac{2}{1+\bar\phi \phi}\bigg]\frac{1}{2\pi}\,\log\frac{M}{\mu}\,,
\label{eq:lagrCP1}
\ee
where $M$ denotes the regulator mass -- upper cut-off in the loop integration -- and $\mu$ denotes the normalization point - the lower cut-off in the loop integration. The second piece comes from from the dispersive loop, Fig.~\ref{fig:CP1Diagrams}b, where two vertices are connected by two propagators,
\be
\Delta \mc L_{\rm dispersive}=\p_\m \bar\phi \p_\m \phi\bigg[-\frac{4\bar\phi \phi}{(1+\bar\phi \phi)^{2}}\bigg]\frac{1}{2\pi}\,\log\frac{M}{\mu}\,.
\label{eq:lagrCP2}
\ee
The sum,
\be
\mc L _{CP}+\Delta \mc L_{\rm tadpole}+\Delta \mc L_{\rm dispersive}=\frac{1}{2\pi}\frac{\p_\m \bar\phi \p_\m \phi}{(1+\bar\phi \phi)^{2}}
\bigg[\frac{4\pi}{g_{0}^{2}}-2\log\frac{M}{\mu}\bigg],
\ee
demonstrates the AF result for the one-loop running of coupling. But we would like break it into tadpole and dispersive part to have an analogy with the instanton derivation. To this end let us 
extract the operator $\p_\m \bar\phi \p_\m \phi \,\bar \phi\phi$, which refers to the scattering amplitude, from \REF{eq:lagrCP}, \REF{eq:lagrCP1}, \REF{eq:lagrCP2}. Then we get
\be
-\frac{1}{\pi}\p_\m \bar\phi \p_\m \phi \bar \phi\,\phi\bigg[\frac{4\pi}{g_{0}^{2}}-4\log\frac{M}{\mu} +2\log\frac{M}{\mu}\bigg]
\ee
in clear analogy with the instanton breaking in \REF{eq:CP1g2Renormalization} into zero and nonzero modes contributions.

Mapping the theory on the sphere $\mathbb{S}^2$, we can establish that the spectra of positive modes around the instanton and the vacuum are given by~\cite{Jevicki:1977yd}
\be
\lambda_\ell ^{\mc I} = \ell(\ell+1)-2, ~~ \ell=2, 3, \dots,
\ee
and
\be
\lambda_\ell ^0 = \ell(\ell+1), ~~ \ell=1,2,\dots,
\ee
with multiplicities
\be
\n_\ell = 2(2\ell+1).
\ee
Furthermore, there are four (not six) zero modes around the instanton background. Therefore, computing the analogue of \REF{eq:FLSchematicNonZero} leads to
\be
\mc R_{CP}=M^4 \f{\prod_{\ell=2}^\Lambda \l ( \lambda_\ell^{\mc I}+M^2\r ) ^{2\ell+1}}
{\prod_{\ell=1}^\Lambda\l ( \lambda_\ell^{0}+M^2\r ) ^{2\ell+1}} = 
M^4 M^{-6} \prod_{\ell=1}^\infty \l ( \f{\lambda_\ell^{\mc I}+M^2}
{\lambda_\ell^{0}+M^2} \r ) ^{2\ell+1}.
\ee
Now using that
\be
(2\ell+1) \log \l [ \ell(\ell+1) + a \r ] \underset{\ell \to \infty}{=} \dots + \f{2a}{\ell}+\dots,
\ee
we conclude
\be
\mc R_{CP}= M^4 M^{-6} M^4 = M^2,
\label{eq63}
\ee
consistent with \REF{eq:CP1g2Renormalization}.

\section{The hard cutoff regularization on a sphere\label{app:CutoffSphere}}

In this case in order to find the effective action \REF{eq:Gamma1LoopPV}, we simply truncate the product at a certain large value $\ell = \Lambda$ and omit the contribution from regulators
\be
\label{eq:Gamma1Loop}
\G_d[\phi_c] = S_d[\phi_c] + \f{1}{2} \sum^\Lambda_{\ell=0} \m_\ell(d) \l [ \log \bar \lambda_{\ell} (d) + (N-1) \log \lambda_{\ell} (d) \r ].
\ee
To find the divergent (in $\Lambda$) part, we expand the sum above for large $\ell$. Let us focus on the case $N=1$. Evaluating the sum in \REF{eq:Gamma1Loop} leads to the following expressions in $d=3,4,6$, where we write explicitly only the $\log \Lambda$ terms, giving other coefficients schematically,
\bea
\label{eq:G3Unsubtrcted}
\G_3[\phi_c] & = & S_3[\phi_0]+\sum_{k=1}^3 a_k\Lambda^k-\f{g}{48}\Lambda\phi_c^4+\f{1}{3} \log \Lambda, \\
\label{eq:G4Unsubtrcted}
\G_4[\phi_c] & = & S_4[\phi_0]+\sum_{k=1}^4 a_k\Lambda^k-\f{g}{12}\Lambda^2\phi_c^2
-\f{g}{8}\Lambda\phi_c^2+ \l( \f{1}{3} -\f{g^2\phi_c^4}{48} \r )\log \Lambda, \\
\label{eq:G6Unsubtrcted}
\G_6[\phi_c] & = & S_6[\phi_0]+\sum_{k=1}^4 a_k\Lambda^k+\dots+ \l (\f{3}{10} + \f{g^2\phi_c^2}{120}
-\f{g^3\phi_c^3}{360} \r )\log \Lambda.
\eea
Several comments are in order. First, the presence of all powers of $\Lambda$ is an illustration of the fact that the cutoff regularization breaks diff invariance. Otherwise powers of $\Lambda$ would be only $d$, $d-2$, $d-4$ etc.
Second, as is evidenced by different powers of $\phi_c$, there are multiple contributions to the $\log \Lambda$ coefficient. Moreover, in three dimensions (and formally in any odd number of dimensions, although for $d\neq 3$ those theories cannot be formulated perturbatively around a trivial vacuum), this coefficient does not correspond to~$(\phi^2_i)^{3}$, which is in one-two-one correspondence with the vanishing beta function at one-loop order.

\paragraph{In $d=4$} using that
\be
S_4[\phi_c] = \phi_c^2-\f{8\pi^2}{3}\f{g \phi_c^4}{4!} 
\ee
 the coupling should be substituted by
\be
g_R=g\l ( 1 +\f{3g}{16\pi^2} \log \Lambda \r ),
\ee
which correctly reproduces the running. Including also $N-1$ transverse modes, we get
\be
\label{eq:EA4Unsubtracted}
\G_4[\phi_c] = S_4[\phi_c]+ \dots + \l( \f{1}{3} +\f{N-1}{3} -\f{1}{2}\f{g^2\phi_c^4}{4!}- \f{N-1}{18}\f{g^2\phi_c^4}{4!} \r )\log \Lambda,
\ee
leading to
\be
\b_4(g) = - \f{N+8}{3}\f{g^2}{16\pi^2}.
\ee
Imagine that we computed the effective action around the instanton background\,\footnote{\,We neglect for the moment the existence of zero modes. As usual, those get converted into integrals over moduli.} we would obtain
\be
\label{eq:InstantonFluctuationsUnsubtracted}
\G_4[\phi_F^s] = \f{16\pi^2}{g}+N \log \Lambda - \f{N+8}{3} \log \Lambda,
\ee
which does not reproduce the correct running of the coupling. Clearly, the reason for this discrepancy is the second term on the right hand side of \REF{eq:InstantonFluctuationsUnsubtracted}, which didn't come from $g \phi_c^4$ operator. In this case an easy fix is to compute the ratio of determinants around the instanton and the trivial background (relative free energy). Then we would get
\be
\label{eq:EA4DiffLambda}
\G_4[\phi_F^s] - \G_4[0] = \f{16\pi^2}{g} - \f{N+8}{3} \log \Lambda,
\ee
which is correct. The reason is that even though there are other dimension $4$ operators, namely, $R^2$ and 
$\phi^2 R$, that can potentially contribute to the $\log \Lambda$ term in the effective action, only one of them, namely $R^2$, appears in \REF{eq:EA4Unsubtracted}.\footnote{\,The reason is that the relative coefficient between the kinetic term and $\phi^2 R$ is fixed by conformal invariance, therefore, since there is no wave function renormalization at one loop, the coefficient in front of $\phi^2 R$ is not renormalized as well.} However, in general, this procedure will work in a more sophisticated way.

\paragraph{For the $d=6$ case} (only $N=1$ is considered), we see that the operator $\phi^2 R$ does contribute to the $\log \Lambda$ part in \REF{eq:G6Unsubtrcted}. Keeping only the $\log \Lambda$ terms and the bare potential, we have
\be
\G_6[\phi_0] = \Omega_6 \l [ 3 \phi_0^2 \l ( 1 + \f{g^2}{384\pi^3} \log \Lambda\r ) - \f{g\phi_0^3}{3!} \l ( 1+ \f{g^2}{64\pi^3} \log \Lambda \r ) \r ] +\f{3}{10} \log \Lambda
\ee
it is clear that the divergent part in $\phi ^2R$ is removed by introducing a counter term for the wave function renormalization
\be
\phi_R^2 = \l ( 1 +\f{g^2}{120} \f{1}{2\Omega_6} \log \Lambda \r)\phi^2=\l (1 + \f{g^2}{384\pi^3} \log \Lambda \r ) \phi^2.
\ee
Combining it with the $\log \Lambda$ contribution coming from $g^2 \phi_0^3$ term leads to the following redefinition of the coupling
\be
g_R = g \l ( 1+ \f{g^2}{64\pi^3} \log \Lambda \r ) \l ( 1+ \f{g^2}{384\pi^3} \log \Lambda \r )^{-3/2},
\ee
corresponding to the beta function
\be
\label{eq:beta6}
\b_6(g) = -\f{3g^3}{256\pi^3}.
\ee
If we were to compute the relative effective action directly as before, we would get
\be
\label{eq:EA6DiffLambda}
\G_6[\phi_F^s] - \G_6[0] = \f{768\pi^3}{5g^2} \l ( 1 - \f{3g^2}{128\pi^3} \log \Lambda \r ),
\ee
consistent with \REF{eq:beta6}.

\section{The hard cutoff regularization on a plane \label{app:HardCutoff}}

For illustrative purposes, we also include here the computation performed directly in flat space. Let us denote by $f_{n,\ell}(r)$ and $\tilde f_{n,\ell}(r)$ the eigenfunctions corresponding to the positive eigenvalues $\lambda_{n,\ell}$ and $\tilde \lambda_{n,\ell}$
\bea
&&-f''_{n,\lambda}-\f{3}{r} f'_{n,\lambda}+\f{\ell(\ell+2)}{r^2}f_{n,\lambda}-\f{g \phi_F^2}{2} f_{n,\lambda} 
= \lambda_{n,\ell} f_{n,\lambda} \\
&&-\tilde f''_{n,\lambda}-\f{3}{r} \tilde f'_{n,\lambda}+ \f{\ell(\ell+2)}{r^2} \tilde f_{n,\lambda}-\f{g \phi_F^2}{6} \tilde f_{n,\lambda} 
= \tilde \lambda_{n,\ell} \tilde f_{n,\lambda}.
\eea
Introducing new variables
\be
f_{n,\ell}=\f{\c_{n,\ell}}{r^{3/2}}, ~~ \tilde f_{n,\ell}=\f{\tilde\c_{n,\ell}}{r^{3/2}},
\ee
we get an equivalent system of equations
\bea
&&-\c''_{n,\lambda}+\f{\ell(\ell+2)+3/4}{r^2}\c_{n,\lambda}-\f{g \phi_F^2}{2} \c_{n,\lambda} 
= \lambda_{n,\ell} \c_{n,\lambda} \\
&&-\tilde \c''_{n,\lambda}+ \f{\ell(\ell+2)+3/4}{r^2} \tilde \c_{n,\lambda}-\f{g \phi_F^2}{6} \tilde \c_{n,\lambda} 
= \tilde \lambda_{n,\ell} \tilde c_{n,\lambda}.
\eea
As a result we have for the integral
\bea
I_F & = & \int \lambda_-^{-1/2} \, d\m \, e^{-16\pi^2/g}
\prod_{n,\ell,\vec m} \lambda_{n,\ell}^{-1/2} \l ( \prod_{n,\ell,\vec m} \tilde \lambda_{n,\ell}^{-1/2} \r ) ^{N-1} \\
& = &
\int \lambda_-^{-1/2} \, d\m \, e^{-16\pi^2/g}
\prod_{\ell} \det \l ( - \f{d^2}{dr^2} + \f{\ell(\ell+2)+3/4}{r^2}-\f{g \phi_F^2}{2} \r ) ^ {-\n (\ell)/2}\nn \\
&&\hspace{3.5cm} \times \l [ \prod_{\ell} \det \l ( - \f{d^2}{dr^2} + \f{\ell(\ell+2)+3/4}{r^2}-\f{g \phi_F^2}{6} \r ) \r]^{-\n (\ell)(N-1)/2},
\eea
The radial determinant can be computed using the Gelfand-Yaglom method. Namely, using the following formula for the ratio of the two determinants
\be
\f{\det \l[  \dst - \f{d^2}{dr^2} + \f{\ell(\ell+2)+3/4}{r^2}+U_1(r) \r ]}
{\det \l[  \dst - \f{d^2}{dr^2} + \f{\ell(\ell+2)+3/4}{r^2} +U_2(r)\r ]}=
\lim_{R\to \infty}\f{F_1(R)}{F_2(R)},
\ee
with $F_i(r)$ the solutions of
\be
-F_i''(r)+\f{\ell(\ell+2)+3/4}{r^2}F_i(r)+U_i(r)F_i(r)=0,
\ee
with the following boundary conditions $F_1(\eps)=F_2(\eps)=0$ and $F'_1(\eps)=F'_2(\eps)=1$.

For the case at hand, we normalize with respect to the determinant without any potential. Doing that leads to
\be
\mathcal R_N(\ell) \equiv \f{\det \l[  \dst - \f{d^2}{dr^2} + \f{\ell(\ell+2)+3/4}{r^2}-\f{g \phi_F^2}{2} \r ]}
{\det \l[  \dst - \f{d^2}{dr^2} + \f{\ell(\ell+2)+3/4}{r^2} \r ]}=
\f{\ell (\ell-1)}{(\ell+2)(\ell+3)},
\ee
and
\be
\mathcal R (\ell) \equiv  \f{\det \l[  \dst - \f{d^2}{dr^2} + \f{\ell(\ell+2)+3/4}{r^2}-\f{g \phi_F^2}{6} \r ]}
{\det \l[  \dst - \f{d^2}{dr^2} + \f{\ell(\ell+2)+3/4}{r^2} \r ]}=
\f{\ell}{\ell+2},
\ee
allowing to rewrite the integral as
\be
I_F =
\det \l( -\p^2 \r )^{-N/2}\int \lambda_-^{-1/2} \, d\m \, \exp \l [ -\f{16\pi^2}{g} 
- \f{1}{2}\sum_{\ell} \n(\ell) \, \mc \log R_N(\ell) - \f{N-1}{2}\sum_{\ell} \n(\ell) \, \mc \log R_N(\ell) \r ].
\ee
Expanding summands for large $\ell$ as (see~\cite{Monin:2016bwf} for more)
\be
\f{1}{2}\n(\ell) \, \mc \log R_N(\ell) \underset{\ell \to \infty} {=} -3\ell -3 - \f{3}{\ell}+O(\ell^{-2}),
\ee
and
\be
\f{1}{2}\n(\ell) \, \mc \log R(\ell) \underset{\ell \to \infty} {=} -\ell - 1 - \f{1}{3\ell}+O(\ell^{-2}),
\ee
and using the cutoff $L$, we conclude that
\be
I_F =
\det \l( -\p^2 \r )^{-N/2}\int \lambda_-^{-1/2} \, d\m \, \exp \l [ -\f{16\pi^2}{g} + \f{N+8}{3}\log L 
+ b_2 L^2 + b_1 L\r ].
\ee
Thus, we indeed reproduce the running of the coupling
\be
\f{16\pi^2}{g} - \f{N+8}{3}\log \Lambda  = \f{16\pi^2}{g_R}.
\ee

\newpage

\bibliographystyle{utphys}
\bibliography{Fubini}{}

\end{document}